\documentstyle[twoside,fleqn,npb,epsfig]{article}
\input{psfig.tex}
\def\lsim{\mathrel{\rlap{\lower4pt\hbox{\hskip1pt$\sim$}}
    \raise2pt\hbox{$<$}}} 
\def\gsim{\mathrel{\rlap{\lower4pt\hbox{\hskip1pt$\sim$}}
    \raise2pt\hbox{$>$}}} 

\newcommand{\mev}{{\rm Me}\kern-1.pt{\rm V}}
\newcommand{\gev}{{\rm Ge}\kern-1.pt{\rm V}}
\newcommand{\gevsq}{\mbox{$\mathrm{{\rm Ge}\kern-1.pt{\rm V}}^2$}}

\newcommand{\rhoz}{\mbox{$\rho^0$}}

\newcommand{\qsq}{\mbox{$Q^2$}}

\newcommand{\phih}{\mbox{${\phi}_h$}}

\newcommand{\thetah}{\mbox{${\theta}_h$}}

\newcommand{\abst}{\mbox{$|t|$}}

\title{\vskip -20mm \hfill {\normalsize BONN-HE-99-03}\\ \hfill{\normalsize May, 1999}\\*[8mm] Recent Results from Decay-Angle Analyses of {\rhoz} Photoproduction at High Momentum Transfer from ZEUS}

\author{
J.A.Crittenden\thanks{supported by a grant from the Bundesministerium f\"ur Wissenschaft und Forschung of Germany} for the ZEUS collaboration\\ \vskip 1mm
Physikalisches Institut, University of Bonn, Nu{\ss}allee 12, 53115 Bonn, Germany
}

\begin{document}
\begin{abstract}
Recent results from decay-angle analyses of {\rhoz} photoproduction are presented and
discussed in the context of earlier measurements at lower energy and lower
momentum transfer. \vskip -3mm
\end{abstract}
\maketitle

We report recent preliminary results on decay-angle analyses of {\rhoz}
photoproduction at HERA. Measurements of diffractive vector meson production
at HERA have stimulated phenomenological descriptions based on perturbative
QCD, in which the hard scale is provided either by the virtuality of the
exchanged photon or by the mass
of the vector meson. Further theoretical studies have proposed that this scale may be given by the momentum transferred to the proton as well~\cite{gib}. These studies make specific predictions for the helicity 
of the photoproduced {\rhoz}, which we derive from the angular distribution
of the pions produced in its decay. By studying the photoproduction of this
light meson, we address the issue of whether the momentum transferred to the proton can serve as a perturbative scale.

Our analyses are based on two distinct data samples corresponding
to integrated luminosities of 3 pb$^{-1}$ and 27 pb$^{-1}$, the first recorded in 
1995~\cite{ichep96,dis97} and the second recorded in 1996 and
1997~\cite{jerusalem}. During these years HERA operated beams of 27.5~{\gev}
positrons and 820~{\gev} protons. The trigger conditions
required the scattered positron to be detected in a special-purpose
tungsten/scintillator calorimeter located 3~cm from the positron beam axis, 44 meters distant
from the nominal e$^+$p interaction point in the positron-beam flight direction.
The acceptance of this photoproduction tagger is defined by the energy
lost by the positron to the photon which interacts with the proton, thus
determining the center-of-mass energy of the photon-proton system, $W$ ($80<W<120~\gev$).
Since the transverse momentum of the final-state positron is thus required to
be small (${\qsq}<0.01~{\gevsq}$), the transverse momentum of the
{\rhoz} ($p_{\rm t}$) detected in the central detector via its dipion
decay provides an accurate approximation
for the momentum transferred to the proton ($t$) via $t\simeq -p^2_{\rm t}$.
Offline data selection criteria included the reconstruction of exactly two
tracks from the interaction vertex and rejected events with calorimetric energy
deposits in the rear and barrel sections of the calorimeter which were not
associated with the extrapolation of either track. These criteria resulted in
a semi-exclusive topology with a gap between the forward region
of the event and the two tracks of at least two units of rapidity. 
For the decay-angle investigations at high {\abst} presented here, the
squared transverse momentum of the two-pion system relative to the beam axes was
required to exceed 1~{\gevsq} in the 1996/97 event sample.
Information from the forward calorimeter was
used to distinguish elastic from proton-dissociative events in the 1995 data sample, while
for the 1996/97 data at higher {\abst} no attempt was made to exclude
the small elastic contribution.
The 1995 sample consists of about 2000 events and the 1996/97 sample of about
20000 events. 

\begin{figure}[htbp]
\epsfig{file=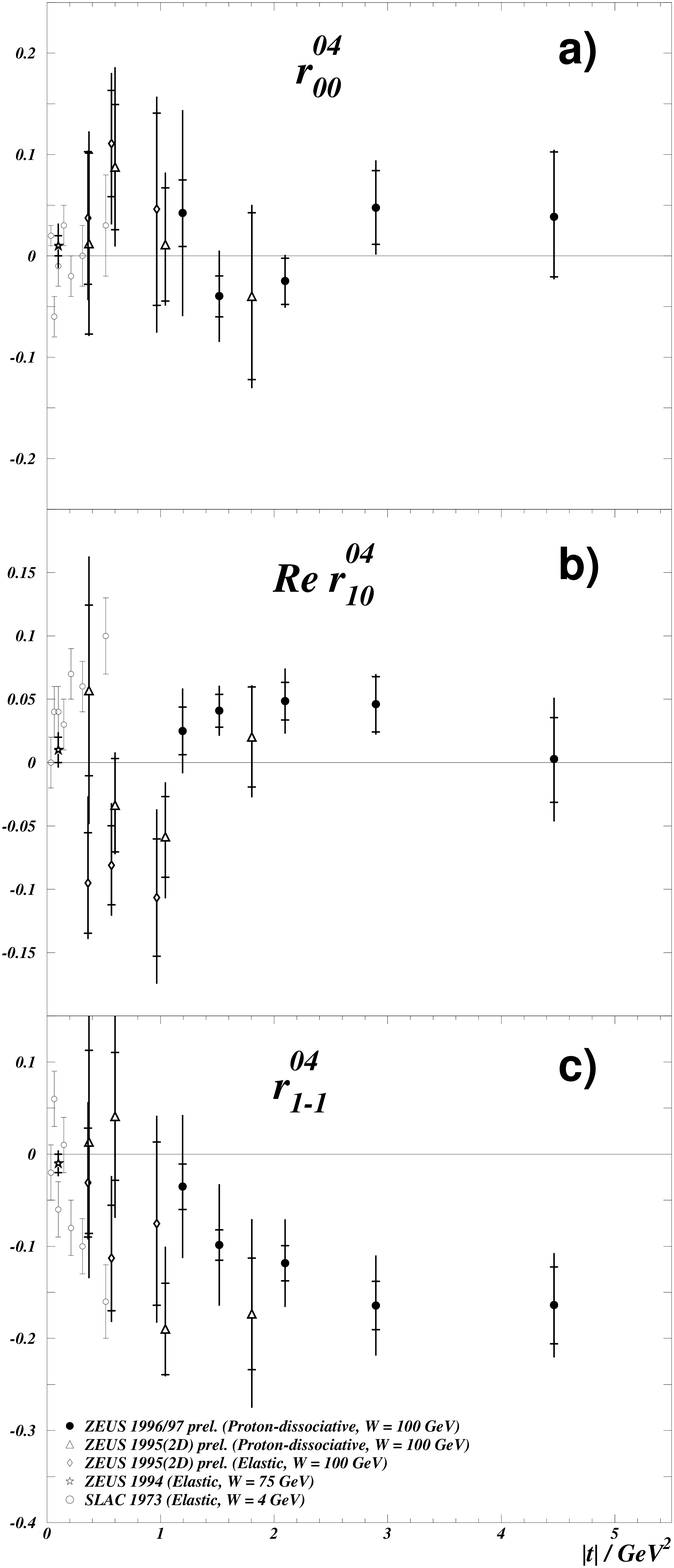, width=5.3cm, bbllx=14, bblly=6, bburx=514, bbury=1650, clip=}
\caption{Measurements of the combinations of matrix elements
  a) $r^{04}_{00}$, c) Re $r^{04}_{10}$, c) $r^{04}_{1-1}$ for the diffractive
  photoproduction of pion pairs. See text for full description}
\label{fig:rij}
\end{figure}

The decay-angle distribution parameterized in terms of combinations
of spin-density matrix elements in the Schilling-Wolf convention~\cite{sw}, 
$r^{04}_{ij}$, is given by 
\begin{eqnarray*}
\begin{array}{ll}{W(\thetah,\phih)}={3\over 4\pi} \biggl[
\frac{1}{2} \left(1-{{r^{04}_{00}}}\right)+\frac{1}{2} 
\left(3{{r^{04}_{00}}}-1\right)\cos^2{\thetah} \\[4mm]
-\sqrt{2}{{{\rm Re}(r^{04}_{10})}} \sin 2\thetah\cos\phih
-{{r^{04}_{1-1}}}\sin^2\thetah\cos{2\phih}
\;\biggr],\end{array}
\end{eqnarray*}
where {\thetah} and {\phih} are the polar and azimuthal angles of the
positively-charged pion relative to the $z$-axis defined as the direction
opposite the final-state proton system in the {\rhoz} rest frame, i.e. the
direction of the {\rhoz} momentum in the photon-proton center-of-mass 
frame. This choice of $z$-axis defines the `$s$-channel helicity frame'.
The origin of the azimuthal angle is defined by the 
{\rhoz} production plane. The three-dimensional distribution has been
averaged over the azimuthal angle between the positron scattering
plane and the {\rhoz} production plane, and thus no longer distinguishes the
photon helicity states $\pm 1$.
Figure~\ref{fig:rij} 
shows the results for the combinations
of matrix elements obtained from a least-squares minimization procedure
in which they served as fit parameters.
The systematic
uncertainties are dominated by the uncertainty in the acceptance corrections.
Slightly different dipion mass ranges were used in the two studies:
\mbox{$0.55<M_{\pi\pi}<1.2~{\gev}$} for the 1995 data and 
\mbox{$0.45<M_{\pi\pi}<1.1~{\gev}$} for the 1996/97 data sample.
The results are compared to the results at lower {\abst} for the elastic
reaction obtained with 9 {\gev}
photons from a backscattered laser beam at SLAC incident on a hydrogen bubble 
chamber~\cite{ballam}. Also shown are the ZEUS 1994 results for elastic {\rhoz}
photoproduction at low {\abst}~\cite{z94rho}. The parameter $r^{04}_{00}$,
which is proportional to the square of the amplitude for producing {\rhoz}
mesons in helicity
state 0, is consistent with zero over the entire range in {\abst}. 
The combination \mbox{Re $r^{04}_{10}$}, which is predominantly sensitive to the interference between
the helicity-conserving amplitude and the single-flip amplitude, shows slight
evidence for a single-flip contribution in both the SLAC data and the
high-{\abst} ZEUS results. A clear indication of a double-flip contribution is
shown by the measurements of $r^{04}_{1-1}$ at high {\abst}, consistent with
the SLAC results. There is no
evidence for any difference between the angular distributions in the elastic
and proton-dissociative production processes.

\begin{figure}[htbp]
\vspace*{-5mm}
\epsfig{file=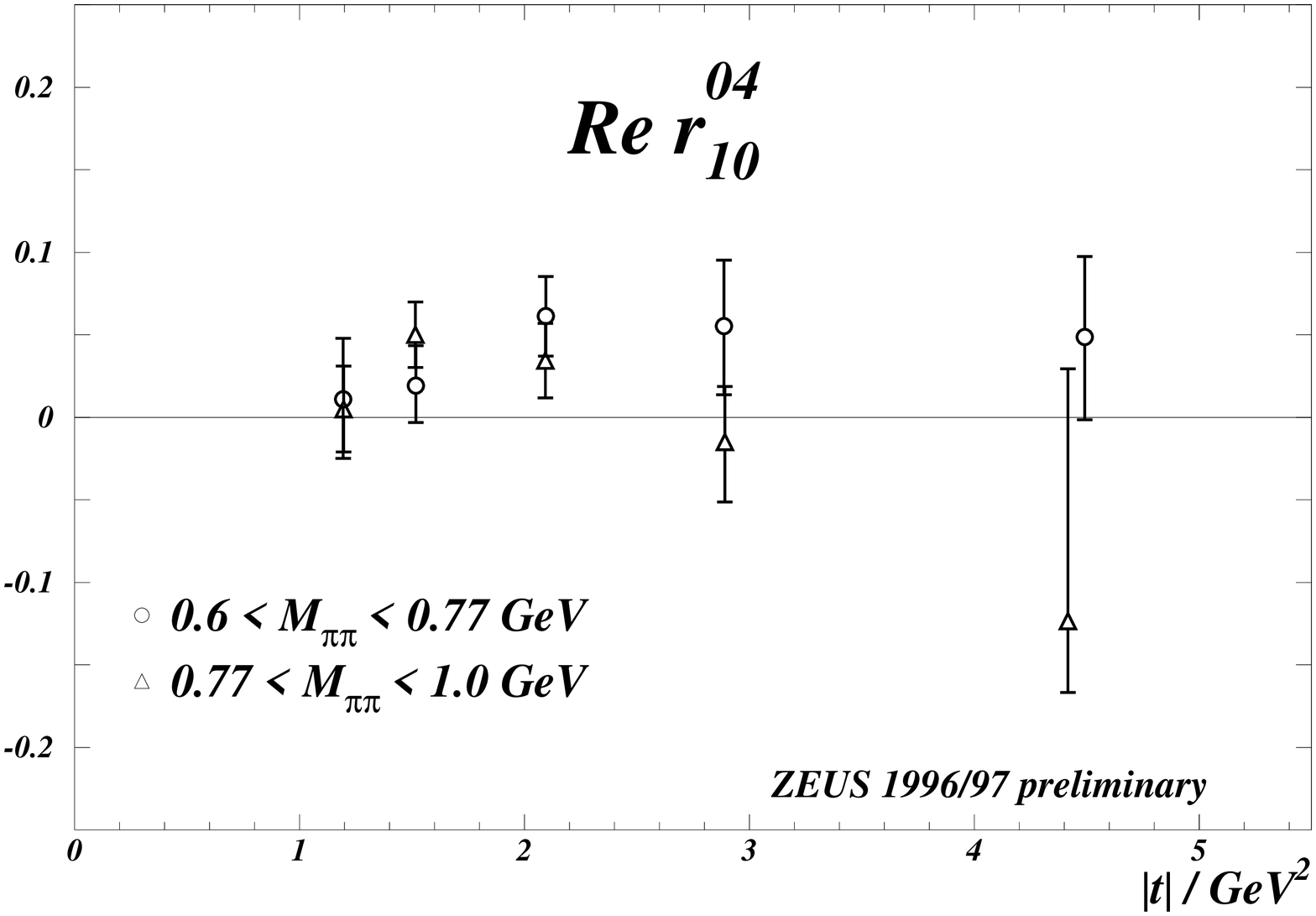, width=5cm, bbllx=11, bblly=8, bburx=514, bbury=514, clip=}
\vspace*{-5mm}
\caption{Results for Re $r^{04}_{10}$ at high {\abst} restricted to the
dipion mass ranges \mbox{$0.6<M_{\pi\pi}<0.77~{\gev}$} and \mbox{$0.77<M_{\pi\pi}<1.0~{\gev}$}\vspace*{-5mm}}
\label{fig:mpi} 
\end{figure}

In order to estimate the effect on the angular distributions of a 
hypothesized dipion background to {\rhoz} decay, the decay-angle 
analysis was repeated for restricted dipion mass ranges above 
(\mbox{$0.77<M_{\pi\pi}<1.0~{\gev}$}) and
below (\mbox{$0.6<M_{\pi\pi}<0.77~{\gev}$}) the nominal value for the {\rhoz} mass. While the above conclusions for
$r^{04}_{00}$ and $r^{04}_{1-1}$ were found to apply also to each mass range separately,
the observed value for \mbox{Re $r^{04}_{10}$} depends significantly 
on the mass range chosen, as shown in Fig.~\ref{fig:mpi}. 
The extraction of the {\rhoz} spin-density
matrix elements from these angular distributions
awaits the understanding of this dependence on the dipion invariant mass.

\begin{figure}[t]
\vspace*{-35mm}
\epsfig{file=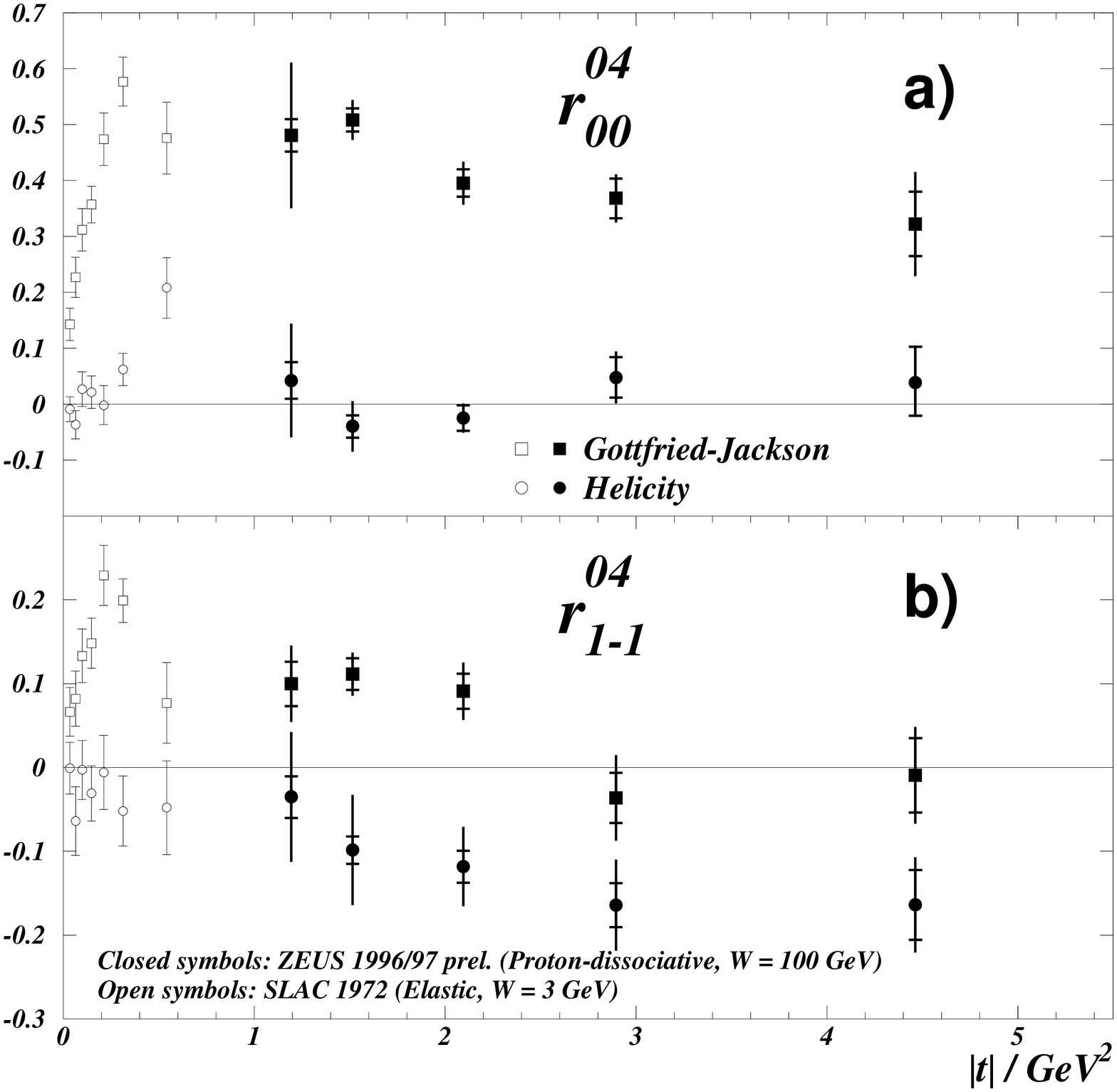, width=5cm, bbllx=10, bblly=12, bburx=514, bbury=1088, clip=}
\vspace*{-5mm}
\caption{Comparison of the combinations of spin-density matrix elements determined in the 
helicity and Gottfried-Jackson frames for a) $r^{04}_{00}$ and b) $r^{04}_{1-1}$\vspace*{-7mm}}
\label{fig:gij}
\end{figure}

The decay-angle analysis was also performed in the Gottfried-Jackson frame,
where the $z$-axis is defined to be the direction of the photon momentum in
the photon-proton center-of-mass system, boosted to the {\rhoz} rest system.
The angle between these two axes approaches $180^\circ$ in the limit ${\abst}\gg
M^2_{\pi\pi}$. For this study the $r^{04}_{ij}$ were determined via
one-dimensional fits to acceptance-corrected angular distributions
in $\cos\theta$ and
$\phi$, and consistency with the two-dimensional fits described above was
verified for the measurements in the helicity frame. Figure~\ref{fig:gij}
shows the extension of the low energy, low {\abst} SLAC data to higher
{\abst}, whereby here the 5~{\gev} photon data~\cite{ballam1} 
from SLAC are used, since
the frame comparison for the 9~{\gev} data was not published. 
Our results confirm the trends observed at low energy in each of the two frames.

In summary, decay-angle analyses of {\rhoz} photoproduction have been
extended to higher values of {\abst} at $W \simeq 100~\gev$. The violation
of $s$-channel helicity conservation observed at lower energy and lower
momentum transfer is confirmed by the new results. More work is needed
to understand the dependence of the extracted spin-density matrix elements
on dipion invariant mass.

\end{document}